\documentclass [12pt]{article}
\usepackage{amsmath,amssymb,cite}

\usepackage{tabularx}
\usepackage[english]{babel}
\usepackage[dvips]{graphicx}
\usepackage{indentfirst}
\usepackage{epsfig}
\numberwithin{equation}{section}

\begin{document}

\title{Wilson RG of Noncommutative $\Phi_{4}^4$}

\author{ Badis Ydri $^{a}$\footnote{Email:ydri@stp.dias.ie,~badis.ydri@univ-annaba.org.},  Rachid Ahmim$^{b}$, Adel Bouchareb $^{a}$ \\
$^{a}$ Institute of Physics, BM Annaba University,\\
BP 12, 23000, Annaba, Algeria.\\
$^{b}$  Department of Physics, El-Oued University,\\
BP 789, 39000, El-Oued, Algeria.
}

\maketitle
\abstract{We present a study of phi-four theory on noncommutative spaces using a combination of the  Wilson renormalization group recursion formula and the solution to the zero dimensional vector/matrix models at large $N$. Three fixed points are identified. The matrix model $\theta=\infty$ fixed point which describes the disordered-to-non-uniform-ordered transition. The Wilson-Fisher fixed point at $\theta=0$ which describes the disordered-to-uniform-ordered transition, and a noncommutative Wilson-Fisher fixed point at a maximum value of $\theta$ which is associated with the transition between non-uniform-order and uniform-order phases.}% Two cases are considered: $i)$ in the limit $\theta\longrightarrow 0$ with $O(N)$ vector symmetry and $ii)$ in the limit $\theta\longrightarrow \infty$ with only two noncommuting directions in the matrix basis at the self-dual point }
%\tableofcontents 
\section{Introduction}

A noncommutative field theory is a non-local field theory in which we replace the ordinary local point-wise multiplication of fields with
 the non-local Moyal-Weyl star product \cite{Groenewold:1946kp,Moyal:1949sk}. This product is intimately related to coherent states \cite{Man'ko:1996xv,Perelomov:1986tf,klauder:1985}, Berezin  quantization \cite{Berezin:1974du} and deformation quantization \cite{Kontsevich:1997vb}. It is also very well understood that the underlying operator/matrix structure of the theory, exhibited by the Weyl map  \cite{weyl:1931}, is the singular most important difference with commutative field theory since it is at the root cause of profound physical differences between the two theories.  We suggest \cite{Alexanian:2000uz} and references therein for elementary and illuminating discussion of the Moyal-Weyl product and other star products and their relations to the Weyl map and coherent states.

Noncommutative field theory is believed to be of importance to physics beyond the standard model and the Hall effect \cite{Douglas:2001ba} and also to quantum gravity and string theory \cite{Connes:1997cr,Seiberg:1999vs}. 

Noncommutative scalar field theories are the most simple, at least conceptually, quantum field theories on noncommutative spaces. Some of the novel quantum properties of noncommutative scalar field theory and scalar phi-four theory are as follows: 
\begin{enumerate}
\item The planar diagrams in a noncommutative  $\phi^4$ are essentially identical to the planar diagrams in the commutative theory as shown originally in \cite{Filk:1996dm}. 
\item As it turns out, even the free noncommutative scalar field is drastically different from its commutative counterpart contrary to widespread believe. For example, it was shown in \cite{Steinacker:2005wj} that the eigenvalues distribution of a free scalar field on a noncommutative space with an arbitrary kinetic term is given by a Wigner semicircle law. This is due to the dominance of planar diagrams which reduce the number of independent contractions contributing to the expectation value $<\phi^{2n}>$ from $2^nn!$ to the number $N_{\rm planar}(2n)$ of planar contractions of a vertex with $2n$ legs. See also \cite{Polychronakos:2013nca,Tekel:2014bta,Nair:2011ux,Tekel:2013vz} for an alternative derivation.

\item More interestingly, it was found in \cite{Minwalla:1999px} that the renormalized one-loop action of a noncommutative $\phi^4$ suffers from an infrared divergence which is obtained when we send either the external momentum or the non-commutativity to zero. This non-analyticity at small momenta or small non-commutativity (IR) which is due to the high energy modes (UV) in virtual loops is termed the UV-IR mixing. 
\item We can control the UV-IR mixing found in  noncommutative $\phi^4$ by modifying the large distance behavior of the free propagator through adding a harmonic oscillator potential to the kinetic term \cite{Grosse:2003aj}.  More precisely, the UV-IR mixing of the theory is implemented precisely in terms of a certain duality symmetry of the new action which connects momenta and positions \cite{Langmann:2002cc}. The corresponding Wilson-Polchinski renormalization group equation \cite{Polchinski:1983gv,Keller:1990ej} of the theory can then be solved in terms of ribbon graphs drawn on Riemann surfaces.  Renormalization of  noncommutative $\phi^4$ along these lines was studied for example in \cite{Chepelev:1999tt,Chepelev:2000hm,Grosse:2003aj,Grosse:2004yu,Grosse:2003nw,Rivasseau:2005bh,Gurau:2005gd}. Other approaches to renormalization of quantum noncommutative $\phi^4$ can be found for example in \cite{Becchi:2002kj,Becchi:2003dg,Gurau:2009ni,Griguolo:2001ez,Sfondrini:2010zm,Gurau:2008vd}.
\item In two-dimensions the existence of a regular solution of the Wilson-Polchinski equation \cite{Polchinski:1983gv} together with the fact that we can scale to zero the coefficient of the harmonic oscillator potential in two dimensions leads to the conclusion that the standard non-commutative $\phi^4$ in two dimensions is renormalizable  \cite{Grosse:2003nw}. In four dimensions, the harmonic oscillator term seems to be essential for the renormalizability of the theory \cite{Grosse:2004yu}.
\item The beta function of noncommutative $\phi^4$ theory at the self-dual point is zero to all orders \cite{Grosse:2004by,Disertori:2006nq,Disertori:2006uy}. This means in particular that the theory is not asymptotically free in the UV since the RG flow of the coupling constant is bounded and thus the theory does not exhibit a Landau ghost, i.e. not trivial. In contrast the commutative $\phi^4$ theory although also asymptotically free exhibits a Landau ghost.  

\item Noncommutative scalar field theory can be non-perturbatively regularized using either fuzzy projective spaces ${\bf CP}^n$ \cite{Balachandran:2001dd} or fuzzy tori ${\bf T}^n$ \cite{Ambjorn:2000cs}. The fuzzy tori are intimately related to a lattice regularization whereas fuzzy projective spaces, and fuzzy spaces \cite{Balachandran:2005ew,O'Connor:2003aj} in general, provide a symmetry-preserving sharp cutoff regularization. By using these regulators noncommutative scalar field theory on a maximally noncommuting space can be rewritten as a matrix model given by the sum of kinetic (Laplacian) and potential terms. The geometry in encoded in the Laplacian in the sense of  \cite{Connes:1994yd,Frohlich:1993es}. 

The case of degenerate noncommutativity is special and leads to a matrix model only in the noncommuting directions. See for example \cite{Grosse:2008df} where it was also shown that renormalizability in this case is reached only by the addition of the doubletrace term $ \int d^Dx (Tr\phi)^2$ to the action.

\item Another matrix regularization of  non-commutative $\phi^4$ can be found  in  \cite{Langmann:2003if,Langmann:2003cg,GraciaBondia:1987kw} where some exact solutions of noncommutative scalar field theory in background magnetic fields are constructed explicitly. Furthermore, in order to obtain these exact solutions matrix model techniques were used extensively and to great efficiency. For a pedagogical introduction to matrix model theory see \cite{Brezin:1977sv,Shimamune:1981qf,DiFrancesco:1993nw,Mehta,eynard,Kawahara:2007eu}. Exact solvability and non-triviality is discussed at great length in \cite{Grosse:2012uv}.

\item A more remarkable property of quantum noncommutative $\phi^4$ is the appearance of a new order in the theory termed the striped phase which was first computed in a one-loop self-consistent Hartree-Fock approximation in the seminal paper \cite{Gubser:2000cd}. For alternative derivations of this order see for example \cite{Chen:2001an,Castorina:2003zv}. It is believed that the perturbative UV-IR mixing is only a manifestation of this more profound property. As it turns out, this order should be called more appropriately a non-uniform ordered phase in contrast with the usual uniform ordered phase of the Ising universality class and it is related to spontaneous breaking of translational invariance. It was numerically  observed in $d=4$ in \cite{Ambjorn:2002nj} and in $d=3$ in \cite{Bietenholz:2004xs,Mejia-Diaz:2014lza} where the Moyal-Weyl space was non-perturbatively regularized by a noncommutative fuzzy torus \cite{Ambjorn:2000cs}. The beautiful result of \cite{Bietenholz:2004xs} shows explicitly that the minimum of the model shifts to a non-zero value of the momentum indicating a non-trivial condensation and hence spontaneous breaking of translational invariance.  

\item Therefore, noncommutative scalar $\phi^4$ enjoys three stable phases: i) disordered (symmetric, one-cut, disk) phase, ii) uniform ordered (Ising, broken, asymmetric one-cut) phase and iii) non-uniform ordered (matrix, stripe, two-cut, annulus) phase. This picture is expected to hold for noncommutative/fuzzy phi-four theory in any dimension, and the three phases are all stable and are expected to meet at a triple point. The non-uniform ordered phase \cite{brazovkii} is a full blown nonperturbative manifestation of the perturbative  UV-IR mixing effect \cite{Minwalla:1999px} which is due to the underlying highly non-local matrix degrees of freedom of the noncommutative scalar field. In \cite{Gubser:2000cd,Chen:2001an}, it is conjectured that the triple point is a Lifshitz point which is a multi-critical point at which a disordered, a homogeneous (uniform) ordered and a spatially modulated (non-uniform) 
ordered phases meet \cite{Hornreich:1975zz}. 
\item In \cite{Chen:2001an} the triple (Lifshitz) point was derived using the Wilson renormalization group approach \cite{Wilson:1973jj}, where it  was also shown that the Wilson-Fisher fixed point of the theory at one-loop suffers from an instability at large non-commutativity. See \cite{Kopietz:2010zz,Bagnuls:2000ae} for a pedagogical introduction to the subject of the functional renormalization group. The Wilson renormalization group recursion formula was also used in \cite{Ferretti:1996tk,Ferretti:1995zn,Nishigaki:1996ts,Hikami:1978ya,Cicuta:1992cz} to study matrix scalar models which, as it turns out, are of great relevance to the limit $\theta\longrightarrow \infty$ of noncommutative scalar field theory \cite{Bietenholz:2004as}.

\item  The phase structure of non-commutative $\phi^4$ in $d=2$ and $d=3$ using as a regulator the fuzzy sphere was studied extensively in \cite{Martin:2004un,GarciaFlores:2009hf,GarciaFlores:2005xc,Panero:2006bx,Medina:2007nv,Das:2007gm,Ydri:2014rea}. It was confirmed that the phase diagram consists of three phases: a disordered phase, a uniform ordered phases and a non-uniform ordered phase which meet at a triple point. In this case it is well established that the transitions from the disordered phase to the non-uniform ordered phase and from the non-uniform ordered phase to the uniform ordered phase originate from the one-cut/two-cut transition in the quartic hermitian matrix model \cite{Brezin:1977sv,Shimamune:1981qf}. The related problem of Monte Carlo simulation of noncommutative $\phi^4$ on the fuzzy disc was considered in \cite{Lizzi:2012xy}.

\item The above phase structure was also confirmed analytically by the multitrace approach  of \cite{O'Connor:2007ea,Saemann:2010bw} which relies on a small kinetic term expansion instead of the usual perturbation theory in which a small interaction potential expansion is performed. This is very reminiscent of the Hopping parameter expansion on the lattice \cite{Montvay:1994cy,Smit:2002ug}. See also \cite{Ydri:2014uaa} for a review and an extension of this method to the noncommutative Moyal-Weyl plane. For an earlier approach see \cite{Steinacker:2005wj} and for a similar more non-perturbative approach see \cite{Polychronakos:2013nca,Tekel:2014bta,Nair:2011ux,Tekel:2013vz}. This technique is expected to capture the matrix transition between disordered and non-uniform ordered phases with arbitrarily increasing accuracy by including more and more terms in the expansion. Capturing the Ising transition, and as a consequence the stripe transition, is more subtle and  is only possible  if we include odd moments in the effective action and do not impose the symmetry $\phi\longrightarrow -\phi$.

\item The multitrace approach in conjunction with the renormalization group approach and/or the Monte Carlo approach could be a very powerful tool in noncommutative scalar field theory. For example, multitrace matrix models are fully diagonalizable, i.e. they depend on $N$ real eigenvalues only,  and thus ergodic problems are absent and the phase structure can be probed quite directly. The phase boundaries, the triple point and the critical exponents can then be computed more easily and more efficiently. Furthermore, multitrace matrix models do not come with a Laplacian, yet one can attach to them an emergent geometry if the uniform ordered phase is sustained. See for example \cite{multitrace_ydri}. Also, it is quite obvious that these multitrace matrix models lend themselves quite naturally to the matrix renormalization group approach of \cite{Brezin:1992yc,Higuchi:1994rv,Higuchi:1993pu,Zinn-Justin:2014wva}.

\end{enumerate}

Among all the approaches discussed above, it is strongly believed that the renormalization group method is the only  non-perturbative coherent framework in which we can fully understand renormalizability and critical behavior of noncommutative scalar field theory in complete analogy with the example of commutative quantum scalar field theory outlined in \cite{ZinnJustin:2002ru}. The Wilson recursion formula, in particular, is the oldest and most simple and  intuitive renormalization group approach which although approximate agrees very well with high temperature expansions \cite{Wilson:1973jj}. In this approximation we perform the usual truncation but also we perform a reduction to zero dimension which allows explicit calculation, or more precisely estimation, of Feynman diagrams.

The goal in the first part of this article is to apply this method to scalar $\phi^4$ field theory at the self-dual point on a degenerate noncommutative spacetime with two strongly noncommuting directions.  See also \cite{Ydri:2013zya}. In the matrix basis this theory becomes, after appropriate non-perturbative definition, an $N\times N$ matrix model where $N$  is a regulator in the noncommutative directions, i.e.  $N$ here has direct connection with noncommutativity itself. More precisely, in order to solve the theory we propose to employ, following \cite{Ferretti:1996tk,Ferretti:1995zn,Nishigaki:1996ts}, a combination of 
\begin{itemize}
\item $i)$ the  Wilson  approximate renormalization group recursion formula \\ and 
\item
$ii)$ the solution to the zero dimensional large $N$ counting problem given in this case by the Penner matrix model which can be turned into  a multitrace  matrix model for large values of $\theta$. 
\end{itemize}
As discussed neatly in  \cite{Ferretti:1996tk} the virtue and power of combining these two methods lies in the crucial fact that all leading Feynman diagrams in $1/N$ will be counted correctly in this scheme including the so-called "setting sun" diagrams. As it turns out the recursion formula can also be integrated explicitly in the large $N$ limit which  in itself is a very desirable property.

In the second part of this article a non perturbative study of the Ising universality class fixed point in noncommutative $O(N)$ model is carried out using  precisely a combination of the above two methods. See also \cite{Ydri:2012nw}. It is found that the Wilson-Fisher fixed point makes good sense only for sufficiently small values of $ \theta$ up to a certain maximal noncommutativity. This fixed point describes the transition from the disordered phase to the uniform ordered phase in the same way that the matrix model fixed point, obtained in the first model, describes the transition from the one-cut (disordered) phase to the two-cut (non-uniform ordered, stripe) phase. 

Another fixed point termed the noncommutative  Wilson-Fisher fixed point is identified in this case. It interpolates between the commutative Wilson-Fisher fixed point of the Ising universality class which is found to lie at zero value of the critical coupling constant $a_*$ of the zero dimensional reduction of the theory and a novel strongly interacting fixed point which lies at  infinite value of $a_*$ corresponding to maximal noncommutativity. This is identified with the transition between non-uniform and uniform orders.

This article is organized as follows:
\begin{enumerate}
\item The $\theta=\infty$ Fixed Point in Self-Dual Degenerate Noncommutative $\Phi^4$.
\begin{itemize}
\item Degenerate Noncommutativity.
\item Wilson RG Recursion Formula.
\item The Zero-Dimensional Matrix Model.
\item $1/\theta$ Correction.
\item $\theta=\infty$ Fixed Point and Critical Exponents.
\item On the Wave Function Renormalization.
\end{itemize}
\item The $\theta=0$ Fixed Point in Noncommutative $O(N)$ Sigma Model.
\begin{itemize}
\item Maximally Noncommuting $O(N)$ Sigma Model.
\item The Noncommutative Wilson-Fisher Fixed Point.
\end{itemize}
\end{enumerate}

\section{The $\theta=\infty$ Fixed Point in Self-Dual Degenerate Noncommutative $\Phi^4$}
\subsection{Degenerate Noncommutativity}
We are interested in  phi-four theory on a degenerate noncommutative Moyal-Weyl space ${\bf R}^{d}_{\theta}={\bf R}^D\times {\bf R}^2_{\theta}$  with a harmonic osicllator term is give by the action

\begin{eqnarray}
S&=&\int d^{d}x~\bigg[{\Phi}\bigg(-\frac{1}{2}{\partial}_i^2+\frac{1}{2}\Omega^2\tilde{x}_i^2-\frac{1}{2}{\partial}_{\mu}^2+\frac{m^2}{2}\bigg){\Phi}+ g {\Phi}_*^4\bigg].\nonumber\\
\end{eqnarray}
In terms of operators this reads
\begin{eqnarray}
S
&=&\nu_2 \int d^{D}x~ Tr_{\cal H}\bigg[\hat{\Phi}\bigg(-\frac{1}{2}\hat{\partial}_i^2+\frac{1}{2}\Omega^2\tilde{x}_i^2-\frac{1}{2}{\partial}_{\mu}^2+\frac{m^2}{2}\bigg)\hat{\Phi}+ g \hat{\Phi}^4\bigg].
\end{eqnarray}
The index $i$ runs over the noncommuting directions while the index $\mu$ runs over the commuting directions. The Planck volume $\nu_2$ is defined by $\nu_2=2\pi\theta$ where $\theta$ is the noncommutativity parameter and $\tilde{x}_i=2(\theta^{-1})_{ij}x_j$. The parameters of the model are the mass $m^2$, the quartic coupling $g$, the harmonic oscillator parameter $\Omega^2$. We can expand the scalar fields in the  Landau basis $\{\hat{\phi}_{m,n}\}$ as (with $x$ standing for  commuting coordinates)
\begin{eqnarray}
\hat{\Phi}=\frac{1}{\sqrt{\nu_2}}\sum_{m,n=1}^{\infty}M_{mn}(x)\hat{\phi}_{m,n}.
\end{eqnarray}
Furthermore, by introducing a matrix regularization we obtain the action (with $g=\nu_2u/N$)

%\begin{eqnarray}
%S&=&\int d^Dx Tr_N\bigg[\frac{1}{2}(\partial_{\mu}M)^2+\frac{1}{2}m^2M^2+\frac{u}{N}M^4+a\bigg(2E M^2+\sqrt{\omega}(\Gamma^+M\Gamma M+M\Gamma^+M\Gamma)\bigg)\bigg].\nonumber\\\label{th}
%\end{eqnarray}
\begin{eqnarray}
S&=&\int d^Dx Tr_N\bigg[\frac{1}{2}(\partial_{\mu}M)^2+\frac{1}{2}m^2M^2+\frac{u}{N}M^4+a\bigg(E M^2+\sqrt{\omega}\Gamma^+M\Gamma M\bigg)\bigg].\nonumber\\\label{th}
\end{eqnarray}
The coupling constants of the theory are  the mass $m^2$, the quartic coupling constant $u/N$, the noncommutativity parameter $\theta$ and the harmonic oscillator parameter $\Omega^2$. The parameters $a$ and $\sqrt{\omega}$ are defined by 
\begin{eqnarray}
a=2\frac{\Omega^2+1}{\theta}~,~\sqrt{\omega}=\frac{\Omega^2-1}{\Omega^2+1}.
\end{eqnarray}
The external sources  $E$ and $\Gamma$ are the matrices given by
\begin{eqnarray}
(\Gamma)_{lm}=\sqrt{m-1}{\delta}_{lm-1}~,~(E)_{lm}=(l-\frac{1}{2}){\delta}_{lm}.
\end{eqnarray}
At the self-dual point we have  $\Omega^2=1$ and thus the theory becomes
 \begin{eqnarray}
S&=&\int d^Dx Tr_N\bigg[\frac{1}{2}(\partial_{\mu}M)^2+\frac{1}{2}m^2M^2+\frac{u}{N}M^4+aE M^2\bigg].
\end{eqnarray}
\subsection{Wilson RG Recursion Formula}
The Wilson renormalization group approach consists in general in the three main steps: $1)$ Integration, $2)$ Rescaling and $3)$ Normalization. In our case here we will supplement the first step of integration with two approximations $a)$ Truncation and $b)$ Wilson Recursion formula. 

\paragraph{Integration:} We start by decomposing  the $N\times N$ matrix $M$ into an $N\times N$ background matrix $B$ and an $N\times N$ fluctuation matrix $F$, viz $M=B+F$. The background $B$ contains slow modes, i.e. modes with momenta less or equal than $\rho\Lambda$ while the fluctuation $F$  contains fast modes, i.e. modes with momenta  larger  than $\rho\Lambda$ where $0<\rho<1$. The integration step involves performing the path integral over the fluctuation $F$ to obtain an effective path integral over the background $B$ alone. We find

\begin{eqnarray}
Z
&=&\int dB~\exp\big(-S[B]-\Delta S(B)\big).
\end{eqnarray}
%The main goal is to compute the effective action $\Delta S(\tilde{M})$ which contains corrections to the operators already present in the original action $S[\tilde{M}]$ together with all possible effective interactions generated by the integration process. 
An exact formula for $\Delta S(B)$ up to the fourth power in the field $B$ is given by the cumulant expansion %(\ref{mainresult}).
%\begin{eqnarray}
%\Delta S(B) &=&\sum_{i}\Delta S_i(B).\label{mainresult0}
%\end{eqnarray}
\begin{eqnarray}
\Delta S(B)&=&4\frac{u}{N}\int d^Dx <Tr_NB^2F^2(x)>_{\rm co}\nonumber\\
&-&8\frac{u^2}{N^2}\int d^Dx\int d^Dy <Tr_NB F^3(x).Tr_NBF^3(y)>_{\rm co}\nonumber\\
&-&8\frac{u^2}{N^2}\int d^Dx\int d^Dy <Tr_NB^2F^2(x).Tr_NB^2F^2(y)>_{\rm co}\nonumber\\
&+&32 \frac{u^3}{N^3}\int d^Dx\int d^Dy \int d^D z <Tr_NB F^3(x).Tr_NB F^3(y).Tr_NB^2F^2(z)>_{\rm co}\nonumber\\
&-&\frac{32}{3} \frac{u^4}{N^4}\int d^Dx\int d^D y\int d^Dz\int d^Dw<Tr_NB F^3(x).Tr_NB F^3(y).Tr_NBF^3(z).Tr_NB F^3(w)>_{\rm co}.\nonumber\\\label{mainresult0}
\end{eqnarray}
The contribution in the $i$th line will be denoted $\Delta S_i(B)$ in the following.
%A derivation of this fundamental result is given in the second appendix. 
The notation "${\rm co}$" stands for the connected component. The first and second terms yield correction to the mass parameter whereas the last three terms yield correction to the quartic coupling constant. The wave function renormalization is obtained from the expansion around $p^2=0$ of the second term which is the most difficult contribution to calculate.

The formula (\ref{mainresult0}) is still very complicated. To simplify it and to get explicit equations we employ the so-called Wilson truncation and Wilson recursion formula. This is usually thought of as part of the integration step. Wilson truncation means that we calculate quantum corrections to only those terms which appear in the original action. Wilson recursion formula is completely equivalent to the use in perturbation theory of the Polyakov-Wilson rules   given by the following two rules: 
\begin{itemize}
\item $1)$ We replace every internal propagator  $1/(k^2+\mu^2)$ by $1/(\Lambda^2+m^2)$\\
 and 
\item $2)$ We replace every momentum loop integral $\int_{\rho\Lambda}^{\Lambda} d^Dk/(2\pi)^D$ by another constant $v_D=\Lambda^D\hat{v}_D$ where $\hat{v}_D$ is given by 
\begin{eqnarray}
\hat{v}_D=\frac{2(1-\rho^D)}{D}\frac{1}{(4\pi)^{{D}/{2}}}\frac{1}{\Gamma({D}/{2})}.
\end{eqnarray}
\end{itemize}
This is a very long and tedious calculation. The results, in the limit $\theta\longrightarrow\infty$, are as follows.

\begin{figure}[htbp]
\centering
%\begin{center}
\includegraphics[width=15cm,angle=0]{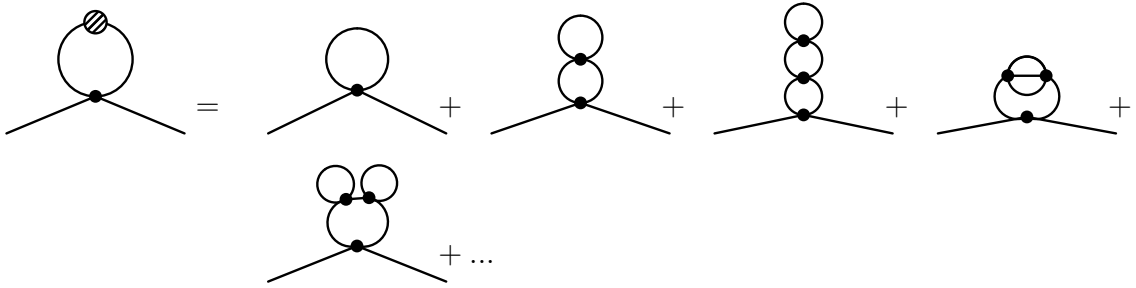}
\caption{Feynman digrams contributing to the renormalization of the mass parameter and the harmonic oscillator coupling constant.}\label{figure1}
%\end{center}
\end{figure}

\begin{figure}[htbp]
\centering
%\begin{center}
\includegraphics[width=15cm,angle=0]{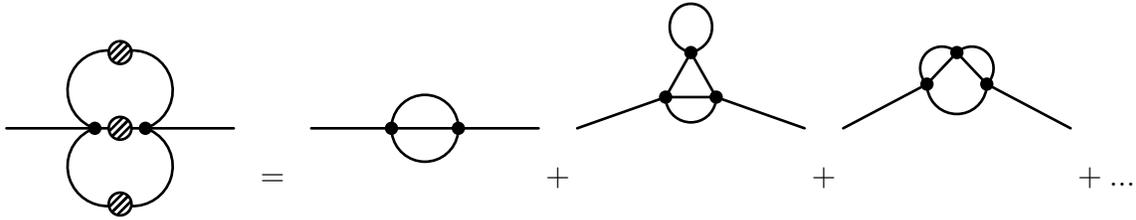}
\caption{Feynman digrams contributing to the wave function renormalization (linear term in $p^2$) and also to the renormalization of the mass parameter and the harmonic oscillator coupling constant (the $p^2=0$ term).}\label{figure2}
%\end{center}
\end{figure}

Quantum corrections to the mass parameter $m^2$ and to the harmonic oscillator coupling constant $a$ are obtained from the first term of (\ref{mainresult0}) and also from the second term  of (\ref{mainresult0}) evaluated at $k^2=0$. The corresponding Feynman diagrams are shown on figures (\ref{figure1}) and (\ref{figure2}). We get
\begin{eqnarray}
S_{{\rm mass}+{\rm harmonic}~{\rm oscillator}}&=&\int d^Dx Tr_N\bigg[\frac{1}{2}m^2B^2+aE B^2\bigg]+\Delta S_1(B)+\Delta S_2(B)|_{k^2=0}\nonumber\\
&=&\int d^Dx Tr_N\bigg[\frac{1}{2}\big(m^2+\Delta m_0^2+a\Delta m_1^2+O(a^2)\big)B^2+a\big(\Delta a_0+a\Delta a_1+O(a^2)\big)E B^2\bigg].\nonumber\\
\end{eqnarray}
The wave function renormalization is also obtained from the  $2$nd term of (\ref{mainresult0}) and as a consequence the relevant Feynman diagrams are still given by those shown on figure (\ref{figure2}). More precisely we need, as before, to expand these diagrams around $k^2=0$ but retain now the linear term in $k^2$ which is very difficult to do explicitly. Our estimation of the coefficient of $k^2$, motivated by dimensional consideration, is obtained by the approximation of  \cite{Ferretti:1995zn,Nishigaki:1996ts}. Explicitly we have the ($3$rd) rule:
\begin{itemize}
\item $3)$ We approximate the first derivative of the propagator with respect to the external momentum $k^2$ by the multiplication with the given propagator as follows
 \begin{eqnarray}
k^2\big[\frac{d}{dk^2}(...)\big]_{k^2=0}&=&-k^2\big[\frac{1}{(k+...)^2+...}(...)\big]_{k^2=0}.\label{ferretti}
\end{eqnarray}
\end{itemize}
We get then
\begin{eqnarray}
S_{{\rm kinetic}}+\partial_{k^2}\Delta S_2|_{k^2=0}=\big[\frac{1}{2}+Z+a\Delta Z+O(a^2)\big]\int d^Dx Tr_N(\partial_{\mu}B)^2.
\end{eqnarray}
The renormalization of the quartic coupling is obtained from the last three terms of  equation (\ref{mainresult0}), i.e. from $\Delta S_3$, $\Delta S_4$ and $\Delta S_5$, and is given explicitly
\begin{eqnarray}
S_{\rm interaction}+\Delta S_3|_{k^2=0}+\Delta S_4|_{k^2=0}+\Delta S_5|_{k^2=0}
&=&\frac{u}{N}\big(1+\Delta u_0+a\Delta u_1+O(a^2))\int d^Dx Tr_NB^4(x).\nonumber\\
\end{eqnarray}
The quantum corrections $\Delta m_0^2$, $\Delta m_1^2$, $\Delta a_0$, $\Delta a_1$, $Z$, $\Delta Z$, $\Delta u_0$ and $\Delta u_1$ will be given explicitly in the next section.
\paragraph{Scaling and Normalization:} By performing the second step of the Wilson renormalization group approach, i.e. by scaling momenta as $p\longrightarrow p/\rho$ so that the cutoff returns to its original value $\Lambda$ and the third and final step of the Wilson renormalization group approach consisting in rescaling the field in such a way that  the kinetic term is brought to its canonical form we obtain the effective action

\begin{eqnarray}
 S+\Delta S&=&\frac{1}{2}\int d^Dx Tr_N(\partial_{\mu}B^{'})^2+\frac{m^{'2}}{2}\int d^Dx Tr_NB^{'2}+a^{'}\int d^Dx Tr_NE B^{'2}\nonumber\\
&+&\frac{u^{'}}{N}\int d^Dx Tr_N B^{'4}.
\end{eqnarray}
The renormalized field $B^{'}$ is related to the bare field $B$ as follows. If $\tilde{B}$ and $\tilde{B}^{'}$  are the Fourier transforms of $B$ and $B^{'}$ respectively then
 \begin{eqnarray}
\tilde{B}^{'}(p)=\sqrt{\rho^{2+D}}\sqrt{1+2(Z+a\Delta Z)+O(a^2)}\tilde{B}(\rho p).\label{wave}
\end{eqnarray}
The renormalized mass ${m}^{'2}$, the renormalized quartic coupling constant ${u}^{'}$ and the renormalized  inverse noncommutativity $a^{'}$ are given by (with $\epsilon=4-D$)
\begin{eqnarray}
{m}^{'2}=\rho^{-2}\frac{m^2+\Delta m_0^2+a\Delta m_1^2+O(a^2)}{1+2(Z+a\Delta Z)+O(a^2)}.
\end{eqnarray}
\begin{eqnarray}
a^{'}=\rho^{-2}\frac{a(\Delta a_0+a\Delta a_1+O(a^2))}{1+2(Z+a\Delta Z)+O(a^2)}.
\end{eqnarray}
\begin{eqnarray}
 {u}^{'}=\rho^{-\epsilon}\frac{u(1+\Delta u_0+a\Delta u_1+O(a^2))}{\big(1+2(Z+a\Delta Z)+O(a^2)\big)^2}.
\end{eqnarray}
The process which led from the bare coupling constants  $m^2$, $a$ and $u$ to the renormalized coupling constants $m^{'2}$, $a^{'2}$ and $u^{'}$ can be repeated an arbitrary number of times. The bare coupling constants will be denoted by $m_{0}^{2}$, $a_{0}$ and $u_{0}$ whereas the the renormalized coupling constants  at the first step of the renormalization group procedure will be denoted by  $m_{1}^{2}$, $a_{1}$ and $u_{1}$. At a generic step $l+1$ of the renormalization group process the renormalized coupling constants  $m_{l+1}^{2}$, $a_{l+1}$ and $u_{l+1}$ are related to their previous values $m_{l}^{2}$, $a_{l}$ and $u_{l}$ by precisely the above renormalization group equations.  We are therefore interested in renormalization group flow in a $3-$dimensional parameter space generated by the mass $m^2$, the quartic coupling constant $u$ and the harmonic oscillator coupling constant (inverse noncommutativity) $a$.

%I quote Wilson and Kogut  as quoted in \cite{Bagnuls:2000ae}: "{ The formal discussion of consequences of the renormalization group works best if one has a
%differential form of the renormalization group transformation. Also, a differential form is useful
%for the investigation of properties of the epsilon expansion to all orders (...) A longer range possibility is
%that one will be able to develop approximate forms of the transformation which can be integrated
%numerically; if so one might be able to solve problems which cannot be solved any other way.}"
%\section{The $\theta=\infty$ Fixed Point}
\subsection{The Zero-Dimensional Matrix Model}  
The explicit calculation of the corrections $\Delta m_0^2$, $Z$ and $\Delta u_0$, at $\theta=\infty$ using the above rules, reduces to the properties of  
the zero-dimensional matrix model
\begin{eqnarray}
V=Tr_N\big[\frac{1}{2}B^2+\frac{g}{N}B^4\big]~,~g=\frac{u v_D}{(m^2)^2}.\label{0m}
\end{eqnarray}
The Schwinger-Dyson identity of this model can be rewritten in terms of the Green's functions $G_2=<Tr_N B^2>/N^2$ and $G_4=<Tr_N B^4>/N^3$ as 
\begin{eqnarray}
1=G_2+4g G_4.
\end{eqnarray}
The model (\ref{0m}) is exactly solvable. The connected $2-$point and $4-$point functions $C_2$ and $C_4$ and the $2-$point and  $4-$point proper vertices $\Gamma_2$ and $\Gamma_4$ of this model are given by (with $r^2=(\sqrt{1+48g}-1)/{24g}$)
\begin{eqnarray}
C_2=G_2~,~\Gamma_2=(C_2)^{-1}~,~G_2=\frac{1}{3}r^2(4-r^2).
\end{eqnarray}
\begin{eqnarray}
C_4=G_4-2(G_2)^2~,~\Gamma_4=-C_4(C_2)^{-4}~,~G_4=r^4(3-r^2).
\end{eqnarray}
Thus, the functions $\Gamma_2(g)$ and $\Gamma_4(g)$ are known non-perturbatively given by 
\begin{eqnarray}
\Gamma_2(g)=\frac{3}{r^2(4-r^2)}=1+8g-80g^2+1664g^3-....
\end{eqnarray}

\begin{eqnarray}
\Gamma_4(g)=\frac{9(1-r^2)(5-2r^2)}{r^4(4-r^2)^4}=4g-32g^2+896g^3+...
\end{eqnarray}
The  Schwinger-Dyson identity of this model can also be rewritten in terms of $\Gamma_2$ and $\Gamma_4$ as
\begin{eqnarray}
\Gamma_2=1+8gG_2-4g\Gamma_4(G_2)^3.
\end{eqnarray}
The corrections $\Delta m_0^2$ and $\Delta u_0$ are found to be given in terms of the $2-$point proper vertex  $\Gamma_2(g)$ and the $4-$point vertex $\Gamma_4(g)$ of the above zero-dimensional matrix model by
\begin{eqnarray}
 \Delta m_0^2=(\Lambda^2+m^2)(\Gamma_2(g)-1).
\end{eqnarray}
\begin{eqnarray}
 \Delta u_0=\frac{\Gamma_4(g)}{4g}-1.
\end{eqnarray}
%Because the model (\ref{0m}) is exactly solvable the functions $\Gamma_2(g)$ and $\Gamma_4(g)$ are known non-perturbatively given by 
%\begin{eqnarray}
%\Gamma_2(g)=\frac{3}{a^2(4-a^2)}=1+8g-80g^2+1664g^3-....
%\end{eqnarray}
%\begin{eqnarray}
%\Gamma_4(g)=\frac{9(1-a^2)(5-2a^2)}{a^4(1-a^2)^4}=4g-32g^2+896g^3+...
%\end{eqnarray}
Similarly, the wave function renormalization $Z$ is found perturbatively to be given by the expansion 
\begin{eqnarray}
Z=8g^2-256g^3+...
\end{eqnarray}
We need now to find a combination of Green's functions and proper vertices of the above zero-dimensional matrix model with an expansion given exactly by $8g^2-256g^3+...$. 
%First, we note that the  Schwinger-Dyson identity of the model (\ref{0m}) can be rewritten in terms of the Green's function $G_2=<Tr_N B^2>/N^2$ as
%\begin{eqnarray}
%\Gamma_2=1+8gG_2-4g\Gamma_4(G_2)^3.
%\end{eqnarray}
From the  Schwinger-Dyson identity of the model (\ref{0m}) we propose that the function $2g\Gamma_4(g)G_2^3(g)$ is the correct guess. Notice the resemblance of  the graphs corresponding to $8gG_2$ and $-4g\Gamma_4G_2^3$ to the graphs associated with the terms  $\Delta S_1$ and $\Delta S_2$ respectively. Indeed, we compute
\begin{eqnarray}
Z&=&2g\Gamma_4(g)G_2^3(g)\nonumber\\
&=&\frac{2g}{3}\frac{r^2(1-r^2)(5-2r^2)}{4-r^2}\nonumber\\
&=&8g^2-256g^3+...
\end{eqnarray}
The renormalization group equations are therefore given by
\begin{eqnarray}
m^{'2}=\rho^{-2}\frac{m^2+2(m^2+\Lambda^2)(\Gamma_2(g)-1)}{1+4g\Gamma_4(g)G_2^3(g)}~,~u^{'}=\rho^{-\epsilon}\frac{\Gamma_4(g)u}{4g\big(1+4g\Gamma_4(g)G_2^3(g)\big)^2}.
\end{eqnarray}
\subsection{$1/\theta$ Correction}
We start with two remarks:
\begin{enumerate}
\item 
The free propagator of this theory is simple given by
\begin{eqnarray}
\Delta_{ij}(k)=\frac{1}{k^2+m^2+a(i+j-1)}.
\end{eqnarray}
In the limit $\theta\longrightarrow \infty$ this propagator behaves as $1/(k^2+m^2)$. In the computation of the effective action we need extensively the sum $\sum_i\Delta_{ij}(k)$. For $\theta=\infty$ this sum is obviously of order $N$. Including also the subleading $1/\theta$ correction this sum  takes then the following form
\begin{eqnarray}
\sum_{i}{\Delta}_{ij}(k)\longrightarrow N {\Delta}_{n_0j}(k)~,~n_0=\frac{N+1}{2}.\label{appr10}
\end{eqnarray}
A straightforward generalization of this result is 

\begin{eqnarray}
 \sum_i{\Delta}^{r_1}_{ij_1}(k_1){\Delta}^{r_2}_{ij_2}(k_2)... \longrightarrow N{\Delta}^{r_1}_{n_0j_1}(k_1){\Delta}^{r_2}_{n_0j_2}(k_2)....\label{appr20}
\end{eqnarray}
%The terms multiplying  $\tilde{M}^2(x)_{ii}$ and $\tilde{M}(x)_{ij}\tilde{M}(x)_{ji}$  still depend on the indices $i$ and $j$ which is the source of the harmonic oscillator renormalization. Recall that the harmonic oscillator term is of the form $\int d^Dx \tilde{M}^2(x)_{ii}(i-1/2)$. The mass correction corresponds to setting these indices equal to some fixed value $n$ whereas the harmonic oscillator correction corresponds to expanding these terms linearly around $n$. 
Again in the spirit of the Wilson contraction we will need to treat the index $j$ in the propagator ${\Delta}_{n_0j}(\Lambda)$ as a continuous  variable and expand the propagator around $j=n$ where $n$ is some index. This actually makes sense since we are assuming that $\theta$ is sufficiently large and thus $a$ is sufficiently small. Similarly to the expansion around $p^2=0$,  only the first two terms in the expansion around $j=n$ are relevant to renormalization here. We choose $n=1/2$ because the harmonic oscillator term is of the form $\int d^Dx B^2(x)_{ii}(i-1/2)$. From these considerations We have then the extra ($4$th) rule

\begin{itemize}
\item $4)$ We expand  ${\Delta}_{n_0j}(\Lambda)$ around $j=n$ as
\begin{eqnarray}
{\Delta}_{n_0j}(c)&=&{\Delta}_{n_0n}(\Lambda)-(j-n)a{\Delta}_{n_0n}^2(\Lambda)+(j-n)^2a^2{\Delta}_{n_0n}^3(\Lambda)+...\label{pro}
\end{eqnarray}
\end{itemize}
\item The explicit calculation of the various quantum corrections using the above rules, for $\theta\neq \infty$, reduces to the properties of  
the zero-dimensional matrix model 
\begin{eqnarray}
V=Tr_N\big[\frac{1}{2}B^2+\frac{g}{N}B^4+\frac{a}{m^2} EM^2\big].
\end{eqnarray}
This model we do not know how to solve exactly, similarly to the $a=0$ model, and thus our results below will be given as perturbative expansions in $g$.
\end{enumerate}
The corrections $\Delta m_0^2$, $Z$ and $\Delta u_0$ are still given by the results of the previous section with the redefinition of $g$ as
\begin{eqnarray}
 g=\frac{uv_D}{(\Lambda^2+m^2+aN)^2}.
\end{eqnarray}
On the other hand, the corrections $\Delta m_1^2$, $\Delta a_0$, $\Delta a_1$, $\Delta Z$ and $\Delta u_1$ are given by the perturbative expansions 

\begin{eqnarray}
 \Delta m_1^2=N(\Gamma_2(g)-\Delta\Gamma_2(g))~,~\Delta\Gamma_2(g)=1-4g+80 g^2-2240g^3+....
\end{eqnarray}
\begin{eqnarray}
 \Delta a_0=\Delta\Gamma_2(g)~,~\Delta a_1=\frac{N}{2(\Lambda^2+m^2)}\big(8g-80 g^2+512 g^3+...\big).
\end{eqnarray}
\begin{eqnarray}
 \Delta Z=\frac{N}{2(\Lambda^2+m^2)}\big(24 g^2-1024 g^3+...\big)..
\end{eqnarray}
\begin{eqnarray}
 \Delta u_1=\frac{N}{2(\Lambda^2+m^2)}\big(-16 g+832 g^2+...\big)..
\end{eqnarray}
\subsection{$\theta=\infty$ Fixed Point and Critical Exponents}
By definition a renormalization group fixed point is a point in the space parameter which is invariant under the renormalization group flow. If we denote the fixed point by $m_*^2$, $a_{*}$ and $u_*$     then we must have 
\begin{eqnarray}
{m}_*^{2}=\rho^{-2}\frac{(m^2+\Delta m_0^2+a\Delta m_1^2+O(a^2))_*}{(1+2(Z+a\Delta Z)+O(a^2))_*}.
\end{eqnarray}
\begin{eqnarray}
a_*=\rho^{-2}\frac{a_*(\Delta a_0+a\Delta a_1+O(a^2))_*}{(1+2(Z+a\Delta Z)+O(a^2))_*}.\label{r*0}
\end{eqnarray}
\begin{eqnarray}
 {u}_*=\rho^{-\epsilon}\frac{u_*(1+\Delta u_0+a\Delta u_1+O(a^2))_*}{(1+2(Z+a\Delta Z)+O(a^2))_*^2}.
\end{eqnarray}
The second equation is new by comparison with the commutative theory. The definition of $g_*$ in terms of $m_*^2$, $a_{*}$ and $u_*$ is obvious. There are possibly several soultions (fixed points) of interest to these renormalization group equations. We will mainly concentrate on the matrix model fixed point corresponding to infinite noncommutativity which is the most obvious solution to equation (\ref{r*0}) given by
\begin{eqnarray}
{a}_{*}=0.
\end{eqnarray}
The remaining two equations reduce then to
\begin{eqnarray}
{m}_*^{2}=\rho^{-2}\frac{(m^2+\Delta m_0^2)_*}{(1+2Z)_*}.\label{mu*10}
\end{eqnarray}
\begin{eqnarray}
 {u}_*=\rho^{-\epsilon}\frac{u_*(1+\Delta u_0)_*}{(1+2Z)_*^2}.\label{u*10}
\end{eqnarray}
Thus this fixed point is fully determined by functions which are known non-perturbatively. An obvious solution to (\ref{u*10}) is $u_*=0$ which corresponds to the usual Gaussian fixed point. By discarding this solution equation (\ref{u*10}) becomes
 \begin{eqnarray}
 1=\rho^{-\epsilon}\frac{\Gamma_4(g_*)}{4g_*(1+2Z(g_*))^2}.\label{u*20}
\end{eqnarray}
The critical value of the mass parameter is obtained from equation  (\ref{mu*10}) as
 \begin{eqnarray}
 \frac{m_*^2}{\Lambda^2}=\frac{\rho^{-2}(\Gamma_2(g_*)-1)}{1+2Z(g_*)-\rho^{-2}\Gamma_2(g_*)}.
\end{eqnarray}
\begin{table}[h]
\centering
\resizebox{7cm}{!}{
\begin{tabular}{|l|c|c|c| }
\hline
$D (d)$ & $g_{*}$  & ${m}_*^2/\Lambda^2$ & ${u}_*/\Lambda^{\epsilon}$\\
\hline 
$1(3)$ & $7.603$ & $-0.935$ & $0.204$\\
\hline
$2(4)$ & $2.282$ & $-0.851$ & $0.854$\\
\hline 
$3(5)$ & $0.409$ & $-0.643$ & $3.527$\\
\hline 
$4(6)$ & $0$ & $0$ & $0$\\
\hline
\end{tabular}
}
\caption{The critical values for $\rho=1/2$.}\label{nta}
\end{table}
The physical region of $g$ is $[0,\infty[$ while the full domain of definition is  $[-1/48,\infty[$. Furthermore, the functions $\Gamma_2$, $\Gamma_4$  and $Z_2=3Z/2g$ depend on $g$ only through  $r=r(g)$  defined by $r^2={2}/({\sqrt{1+48 g}+1})$.  Graphically we observe that the above equation  (\ref{u*10}) admits a non-trivial solution for all dimensions $D=1,2,3,4$ corresponding to $d=3,4,5,6$. The numerical solution for $g_*$, $u_*$ and $m_*^2$ are shown on table (\ref{nta}). There is of course in each dimension the extra Gaussian fixed point as we have discussed. There is only the Gaussian fixed point in $D=4(d=6)$ in this approximation. Also, in our approximation we have checked that there is always a non-trivial fixed point for any value of $\rho$ in the interval $0<\rho<1$. 

In the remainder we will compute the mass critical exponent $\nu$ and the anomalous dimension $\eta$ within this scheme.

\begin{figure}[htbp]
\begin{center}
\includegraphics[width=5.0cm,angle=-90]{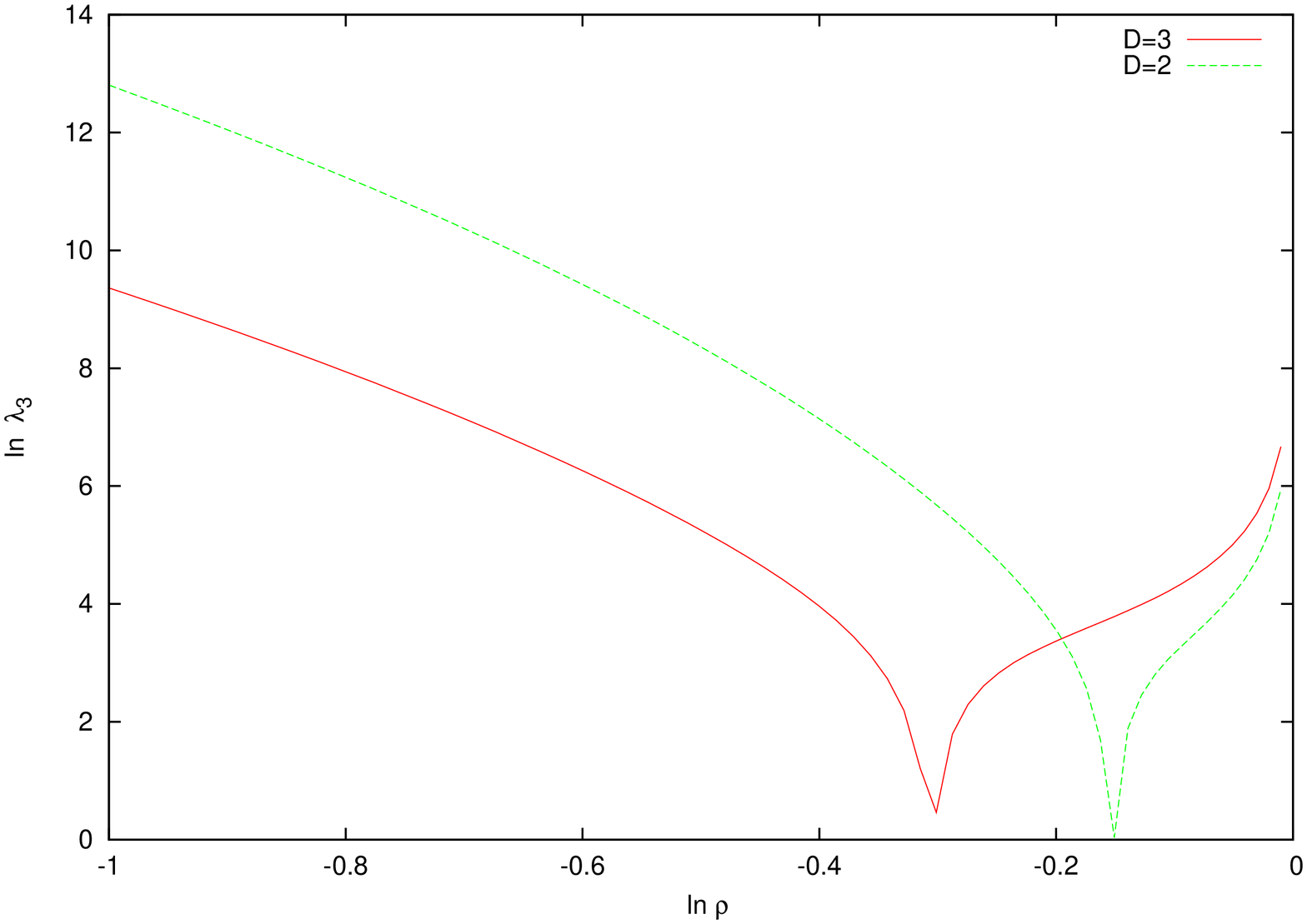}
\caption{The eigenvalue $\lambda_3$ as a function of the dilatation parameter $\rho$.}\label{graphs000}
\end{center}
\end{figure}

%\subsection{Theoretical Digression} 
The computation of the mass critical exponent $\nu$, and other critical exponents, requires linearization of the renormalization above group equations.  These renormalization group equations are of the form
 \begin{eqnarray}
G^{(l+1)}={\cal M}(G^{(l)},\rho).
\end{eqnarray}
The vector of coupling constants $G$ is defined by $G=(G_1,G_2,G_3)$ where  $G_1=m^2$, $G_2=u$ and  $G_3=a$. The linearized renormalization group equations are of the form (with $\delta G=G-G_*$)
 \begin{eqnarray}
\delta G^{(l+1)}={M}(G_*,\rho)\delta G^{(l)}.
\end{eqnarray}
In our problem the matrix $M$ is of the form
\begin{eqnarray}
\left( \begin{array}{ccc}
 M_{11} & M_{12} & M_{13}\\
 M_{21} & M_{22} & M_{23} \\
0 & 0 & M_{33}
\end{array} \right).\label{rgm}
\end{eqnarray}
The eigenvalue in the direction $G_3=a$ is therefore given by
\begin{eqnarray}
{\lambda}_3=M_{33}=\frac{\rho^{-2}(\Delta a_0)_*}{1+2Z_*}.
\end{eqnarray}
This eigenvalue is plotted on figure (\ref{graphs000}) as a function of $\ln \rho$. It looks like that $a$ is an irrelevant coupling constant. 
However, the function $\Delta a_0$ used in the above formula is only known perturbatively and hence this conclusion should be taken with care.

\begin{figure}[htbp]
\begin{center}
\includegraphics[width=5.0cm,angle=-90]{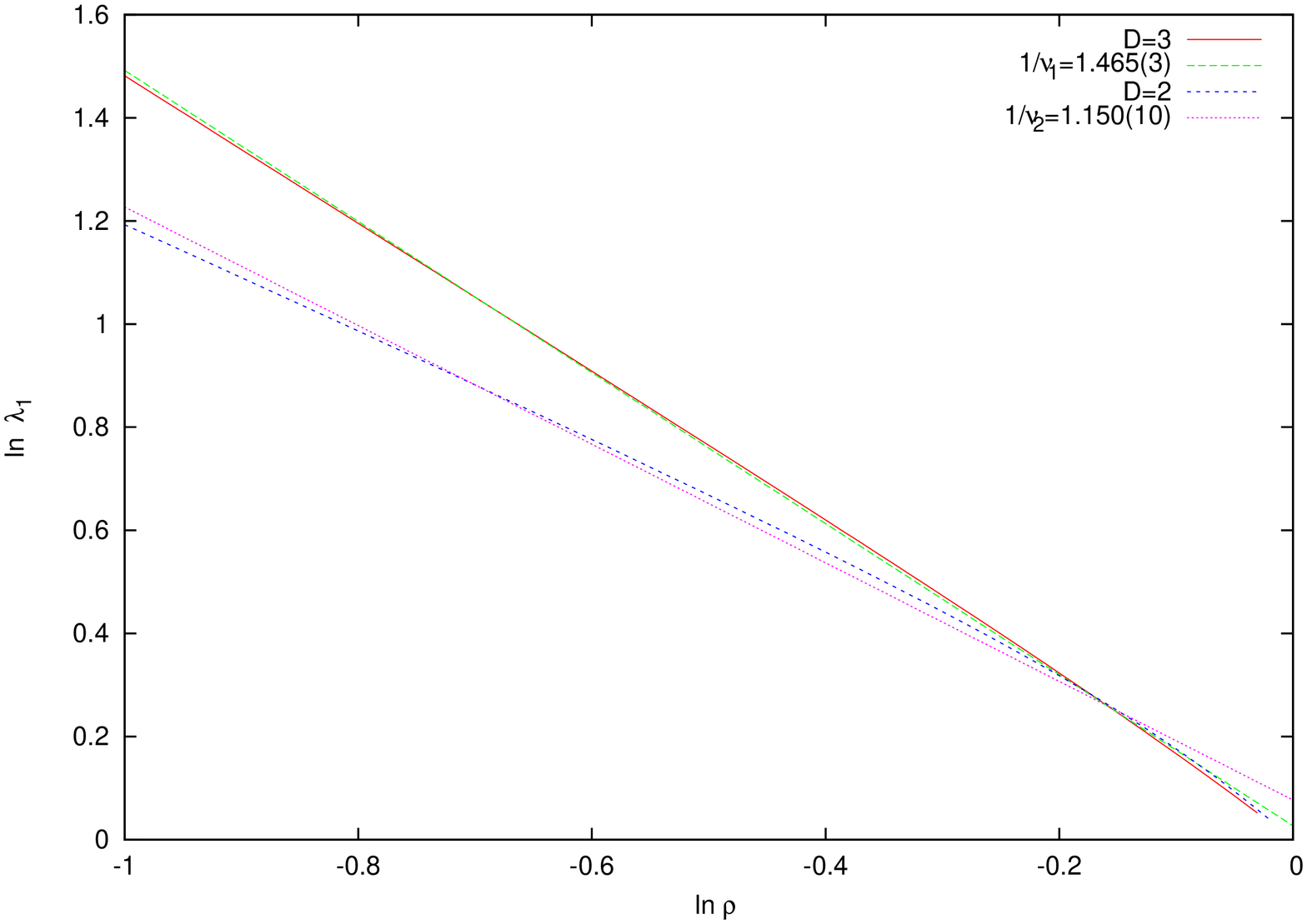}
\includegraphics[width=5.0cm,angle=-90]{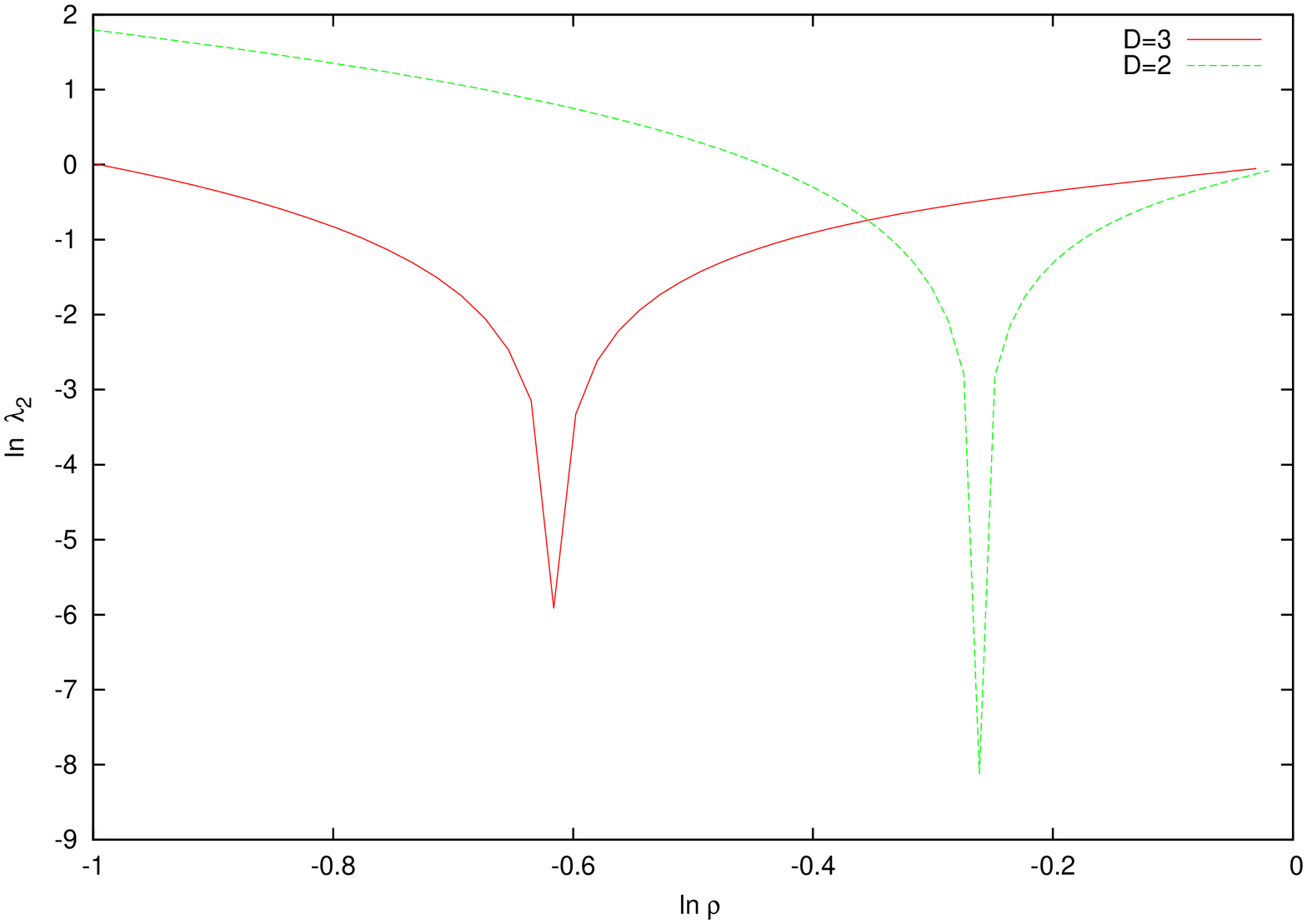}
\caption{The eigenvalues $\lambda_i$ as a function of the dilatation parameter $\rho$.}\label{graphs00}
\end{center}
\end{figure}

The two remaining eigenvalues are determined from the linearized renormalization group equations in the $2-$dimensional space generated by $G_1=m^2$ and $G_2=u$. These are given by
\begin{eqnarray}
\delta m^{'2}=\delta m^2 \big[\frac{m^2}{m^2+\Delta m_{0}^2}\Gamma_2\big]_*+\delta g\bigg[m^2\frac{\Lambda^2+m^2}{m^2+\Delta m_{0}^2}\frac{d\Gamma_2}{dg}-\frac{2m^2}{1+2Z}\frac{dZ}{dg}\bigg]_{*}.
\end{eqnarray}
\begin{eqnarray}
\delta u^{'}=\delta u +\delta g u_*\bigg[\frac{1}{1+\Delta u_{0}}\frac{d}{dg}\big(\frac{\Gamma_4}{4g}\big)-\frac{4}{1+2Z}\frac{dZ}{dg}\bigg]_{*}.
\end{eqnarray}
\begin{eqnarray}
\delta g=\frac{g_*}{u_*}\delta u -\frac{2g_*}{\Lambda^2+m_*^2}\delta m^2.
\end{eqnarray}
As it turns out this problem depends only on functions which are fully known non-perturbatively. The eigenvalues $\lambda_1$ and $\lambda_2$ can be determined from the trace and determinant which are given by
  \begin{eqnarray}
\lambda_1+\lambda_2=M_{11}+M_{22}\equiv {\rm Tr}_2 M~,~\lambda_1\lambda_2=M_{11}M_{22}-M_{12}M_{21}\equiv {\rm det}_2M.
\end{eqnarray}
In other words
 \begin{eqnarray}
\lambda_1=\frac{{\rm Tr}_2 M\pm \sqrt{({\rm Tr}_2M)^2-4{\rm det}_2M}}{2}~,~\lambda_2={\rm Tr}_2 M-\lambda_1.
\end{eqnarray}
The eigenvalues $\lambda_i$ must scale with the dilatation parameter $\rho$ as 
\begin{eqnarray}
\lambda_i(\rho)=\lambda_i(1)\rho^{-y_i}.\label{beha10}
\end{eqnarray}
The exponents $y_i$ are called critical exponents or scaling indices. The mass critical exponent $\nu$ is given by the inverse of the critical exponent of the largest eigenvalue. If $\lambda_1>\lambda_2$ then
\begin{eqnarray}
\nu=1/y_1.
\end{eqnarray}
We found that the solutions $\lambda_1$ and $\lambda_2$ exist for $0.1\leq\rho \leq 1$ for $D=3$ and for $0.35\leq \rho\leq 1$ for $D=2$ with the property $\lambda_1>\lambda_2>0$. The formula (\ref{beha10}) was used then as a crucial test for our numerical calculations. In particular, we have determined by means of this formula the range of the dilatation parameter $\rho$ over which the logarithm  of the eigenvalues scale  linearly with $\ln \rho$.  The eigenvalue $\ln \lambda_1$ was found to be linear over the full range whereas the eigenvalue $\ln \lambda_{2}$ was linear only for $\ln \rho << -1$. In any case, we expect the behavior  (\ref{beha10}) to hold only if the renormalization group steps are sufficiently small so not to alter drastically the infrared physics of the problem.  %Fortunately, it is the eigenvalue $\lambda_1$ which is associated with a relevant scaling field and thus used to define the mass critical exponent as we will discuss shortly. 
Some results are shown on figure (\ref{graphs00}).
%Note that the eigenvalues $\lambda_1$ and $\lambda_2$ are both negative in the regime  $\ln \rho < -1$ which means that the scaling indices $y_1$ and $y_2$ are mildly complex. Furthermore we note that over the  regime $\ln \rho < -1$ the eigenvalue $\lambda_3$ scales also correctly with $\rho$ (figure (\ref{graphs})).
%For $D=2$ the renormalization group steps can be thought of as small and the behavior (\ref{beha1}) holds in the regime $\ln \rho < -1$. As it happens this is the most important case corresponding to $d=4$. 
We find explicitly for $\ln\lambda_1$ the following fits:

\begin{eqnarray}
D=2~,~\ln \lambda_1=-1.150(10) \ln \rho +0.077(5)~,~\frac{1}{\nu_1}=1.150(10).
\end{eqnarray}
\begin{eqnarray}
D=3~,~\ln \lambda_1=-1.465(3) \ln \rho +0.027(2)~,~\frac{1}{\nu_1}=1.465(3).
\end{eqnarray}
%This result is shown on figure (\ref{graphs1}) together with the results for $\ln |\lambda_1|$ for $D=3$ and $D=4$ and their numerical fits. We also  obtain positive slope for $\ln |\lambda_2|$ for $D=3$ and $D=4$ although the range is much more smaller. 
We conclude immediately that the scaling field corresponding to the mass  is relevant since $y_1=1/\nu_1>0$ while the scaling field corresponding to the quartic coupling constant is irrelevant, i.e. $y_1=2/\nu_2<0$, as seen immediately from the behavior of the eigenvalues for $\ln\rho<<1$ on figure (\ref{graphs00}). We skip writing explicitly the corresponding estimate of the coupling constant critical exponent $y_2$.

\begin{figure}[htbp]
\begin{center}
\includegraphics[width=5.0cm,angle=-90]{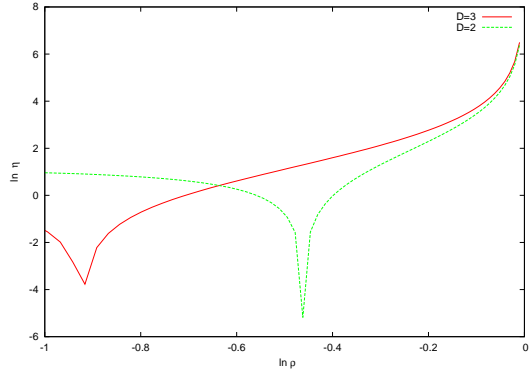}
\caption{The anomalous dimension $\eta$ as a function of the dilatation parameter $\rho$.}\label{graphs0000}
\end{center}
\end{figure}

%\subsection{The Anomalous Dimension $\eta$}
%\paragraph{The Anomalous Dimension $\eta$:}
In order to compute the anomalous dimension $\eta$ we need to go back to the wave function renormalization contained in equation (\ref{wave}) which can be put in the form

\begin{eqnarray}
  \tilde{B}^{'}(p)=\rho^{\frac{2+D-\eta}{2}}\tilde{B}^{}(\rho p).
\end{eqnarray}
The coefficient $\eta$ is called the anomalous dimension. It is given explicitly by

\begin{eqnarray}
  \eta&=&-\frac{\ln(1+2Z_*)}{\ln\rho}\nonumber\\
&=&\frac{\epsilon}{2}-\frac{\ln(\Gamma_4(g_*)/4g_{*})}{2\ln\rho}.
\end{eqnarray}
The results in $D=3$ and $D=2$ are shown on figure (\ref{graphs0000}). We observe that $\eta$ approaches a constant value as $\ln\rho \longrightarrow -1$ for $D=2$.
\subsection{On the Wave Function Renormalization}

The wave function renormalization (\ref{ferretti}) can be improved by replacing the overall minus sign multiplying this equation by the correct coefficient coming from the leading Feynman diagrams. This coefficient is conjectured in \cite{Ydri:2013zya} to be the same for all other subleading Feynman diagrams. The consequences of this change on the fixed point and the critical exponents can be found in \cite{Ydri:2013zya}.

\section{The $\theta=0$ Fixed Point in Noncommutative $O(N)$ Sigma Model}
\subsection{Maximally Noncommuting $O(N)$ Sigma Model}
\begin{figure}[htbp]
\begin{center}
\includegraphics[width=14.0cm,angle=0]{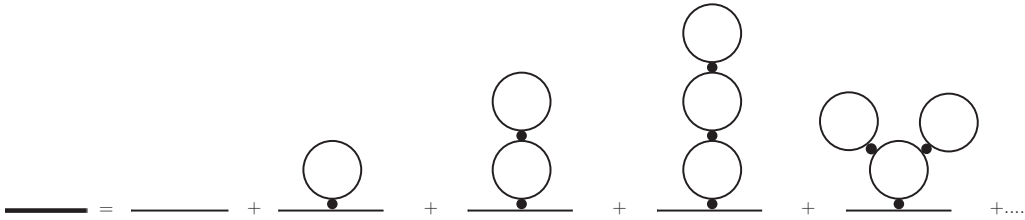}
\caption{The dressed propagator of non-commutative $O(N)$ sigma model.}\label{figure20}
\end{center}
\end{figure}

\begin{figure}[htbp]
\begin{center}
\includegraphics[width=14cm,angle=0]{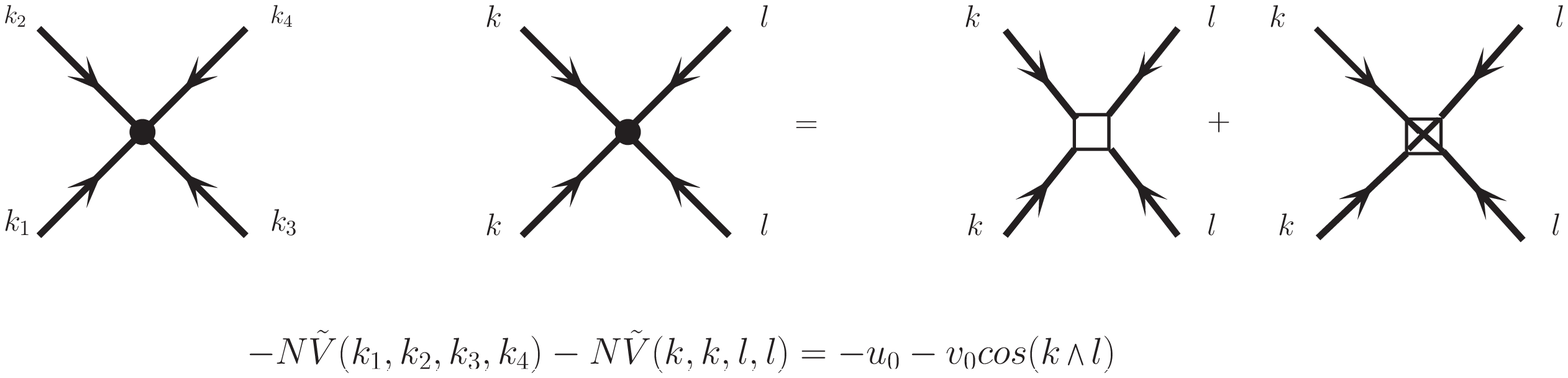}
\caption{The vertex of non-commutative $O(N)$ sigma model.}\label{figure10}
\end{center}
\end{figure}

\begin{figure}[htbp]
\begin{center}
\includegraphics[width=14cm,angle=0]{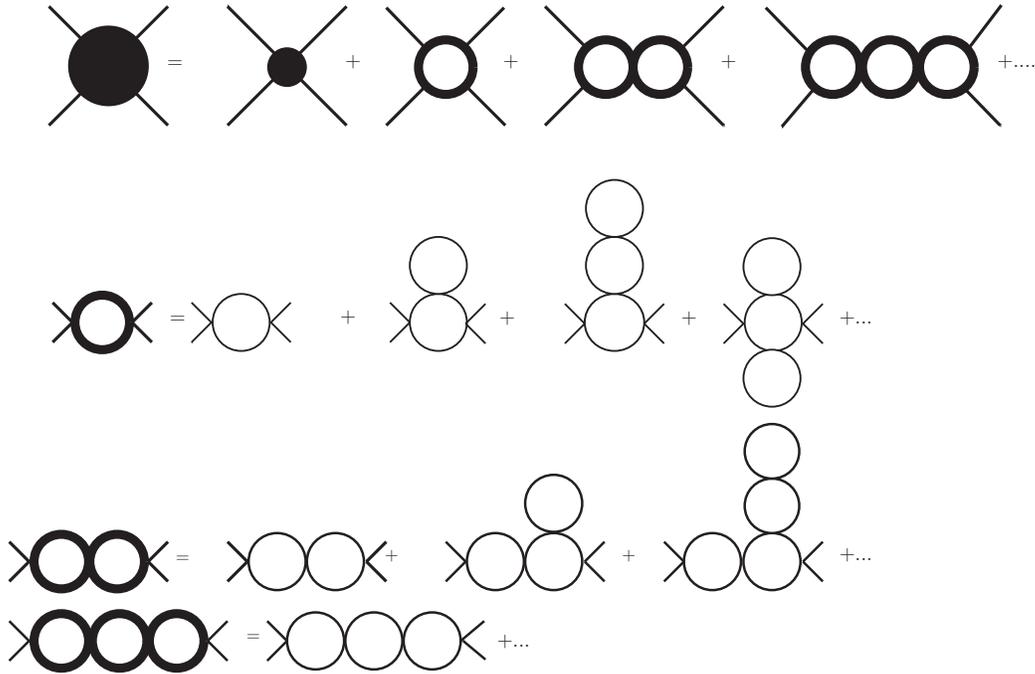}
\caption{The dressed vertex of non-commutative $O(N)$ sigma model.}\label{figure30}
\end{center}
\end{figure}

The action of interest in this section is of the form
\begin{eqnarray}
S=\int d^dx \Phi_a(-{\partial}_i^2+{\mu}^2)\Phi_a+S_{\rm int}+u\int d^dx~(\Phi_a*\Phi_a)^2+v\int d^dx~(\Phi_a*\Phi_b)^2.
\end{eqnarray}
The vertex is given explicitly in momentum space by
\begin{eqnarray}
\tilde{V}(k_1,k_2,k_3,k_4)=u \cos\frac{k_1\wedge k_2}{2}\cos\frac{k_3\wedge k_4}{2}+\frac{v}{2}(e^{\frac{i}{2}(k_1\wedge k_3+k_2\wedge k_4)}+e^{\frac{i}{2}(k_1\wedge k_4+k_2\wedge k_3)}).
\end{eqnarray}
We decompose the fields $\Phi_a(x)$ into backgrounds $\phi_a(x)$ which contain slow modes, i.e. modes with momenta less or equal than $\rho\Lambda$ and fluctuations $f_a(x)$ which contain fast modes, i.e. modes with momenta  larger  than $\rho\Lambda$ where $0<\rho<1$. The partition function is then given by 
\begin{eqnarray}
Z&=&\int d\Phi_a~e^{-S[\Phi]}\nonumber\\%&=&\int d\phi_a~e^{-S[\phi]}\int df_a~e^{-S[f]}e^{-\sigma(\phi,f)}\nonumber\\
&=&\int d\phi_a~e^{-S[\phi]}e^{-\Delta S_{\rm eff}[\phi]}.%<e^{-\sigma(\phi,f)}>\int df_a e^{-S[f]}.
\end{eqnarray}
%We have defined
%\begin{eqnarray}
%<{\cal O}>=\frac{\int df_a~{\cal O}~e^{-S[f]}}{\int df_a e^{-S[f]}}.
%\end{eqnarray}
%The action $\sigma(\phi,f)$ is of the form $\sigma(\phi,f)={\cal M}_1+{\cal M}_2+{\cal M}_3$ where ${\cal M}_n$ contains all the terms which are of order $n$ in the background field $\phi_a$. Clearly the effective action is given by $S_{\rm eff}[\phi]=S[\phi]+\Delta S_{\rm eff}[\phi]$ where $\Delta S_{\rm eff}[\phi]$ is defined by 
%\begin{eqnarray}
%<e^{-\sigma(\phi,f)}>&=&e^{-\Delta S_{\rm eff}[\phi]}.
%\end{eqnarray}
By using the symmetry under $\phi_a\longrightarrow -\phi_a$ and momentum conservation, we compute up to the $4$th order in the slow fields $\phi_a$ the non perturbative expansion 

\begin{eqnarray}
\Delta S_{\rm eff}[\phi]&=&<{\cal M}_2>_{\rm co}-\frac{1}{2}<{\cal M}_1^2>_{\rm co}-\frac{1}{2}<{\cal M}_2^2>_{\rm co}+\frac{1}{2}<{\cal M}_1^2{\cal M}_2>_{\rm co}-\frac{1}{24}<{\cal M}_1^4>_{\rm co}.\label{eff}\nonumber\\
\end{eqnarray}
At this stage we employ the large $N$ limit. After appropriate rescaling, the propagator comes with $1/N$ factor, the vertex comes with a factor of $N$ and the contraction of a vector index yields a factor of $N$. There exists a non-trivial $1/N$ expansion only if $u,v\longrightarrow 0$ when $N\longrightarrow \infty$ such that $u_0=u N$ and $v_0=v N$ is kept fixed. By inspection it is found that all terms %($O(N)$-singlets) 
of the form  $..*\phi_a*..*f_a*..$ are subleading in the large $N$ limit. In other words, we can set in the large $N$ limit ${\cal M}_1,{\cal M}_3\longrightarrow 0$ and 
\begin{eqnarray}
{\cal M}_2&=&2u\int d^dx~\phi_a*\phi_a*f_b*f_b+2v\int d^dx~\phi_a*f_b*\phi_a*f_b.
\end{eqnarray}
As a result the final form of the effective action is given explicitly by the very simple cumulant expansion 
\begin{eqnarray}
\Delta S_{\rm eff}[\phi]&=&<{\cal M}_2>_{\rm co}-\frac{1}{2}<{\cal M}_2^2>_{\rm co}.
\end{eqnarray}
Next we will give the exact solution of the model in the large $N$ limit by computing formally all Feynman diagrams contributing to the $2-$ and the $4-$point function. We start with the correction to the quadratic action given by
\begin{eqnarray}
\Delta S_{\rm quad}[\phi]&=&<{\cal M}_2>_{\rm co}\nonumber\\
&=&\int_{p_1}\tilde{\phi}_a(p_1)\Delta \mu^2(p_1)\tilde{\phi}_a(-p_1),
\end{eqnarray}
where
\begin{eqnarray}
\Delta \mu^2(p_1)=u_0\int_{k_1}\Delta \mu^2(p_1,k_1)+v_0\int_{k_1}\Delta \mu^2(p_1,k_1)\cos k_1\wedge p_1.%e^{-ip_1\wedge k_1}.
\end{eqnarray}
The correction $\Delta \mu^2(p_1,k_1)$ is given by the sum of all bubble graphs shown on figure (\ref{figure20}) with an effective vertex given by a combination of the planar vertex $-u_0$ and the non planar vertex $-v_0\cos p\wedge k$ where $p$ and $k$ are the momenta flowing into the vertex as shown on figure (\ref{figure10}). The result takes the form 

\begin{eqnarray}
\Delta \mu^2(p_1)&=&\delta \mu^2_P+\delta \mu^2_{NP}(p_1).
\end{eqnarray}
\begin{eqnarray}
\delta \mu^2_P&=&u_0\int_{k_1}\frac{1}{k_1^2+\mu^2+\delta \mu^2_P+\delta \mu^2_{NP}(k_1)}.
\end{eqnarray}
\begin{eqnarray}
\delta \mu^2_{NP}(p_1)&=&v_0\int_{k_1}\frac{1}{k_1^2+\mu^2+\delta \mu^2_P+\delta \mu^2_{NP}(k_1)}\cos k_1\wedge p_1.
\end{eqnarray}
Now we discuss the $4-$point function. The full correction to the quartic action in the large $N$ limit is given by
\begin{eqnarray}
\Delta S_{\rm int}[\phi]&=&-\frac{1}{2}<{\cal M}_2^2>_{\rm co}\nonumber\\
&=&\int_{p_1}...\int_{p_4}\tilde{\phi}_a(p_1)\tilde{\phi}_a(p_2)\tilde{\phi}_b(p_3)\tilde{\phi}_b(p_4)~\Delta \tilde{V}(p_1,p_2,p_3,p_4).
\end{eqnarray}
The leading Feynman diagrams in the large $N$ limit contributing to the correction $\Delta \tilde{V}(p_1,..,p_4)$ are shown on figure (\ref{figure30}). For simplicity, we skip writing them explicitly.

The full action: the classical+the complete quantum corrections in the large $N$ limit for the quadratic and quartic terms is therefore given by

\begin{eqnarray}
S_{\rm eff}&=&\int_{p\leq \rho\Lambda}\tilde{\phi}_a(p)\big(p^2+\mu_{\rm eff}^2\big)\tilde{\phi}_a(-p)\nonumber\\
&+&\int_{p_1\leq \rho \Lambda}...\int_{p_4\leq \rho\Lambda}\tilde{\phi}_a(p_1)\tilde{\phi}_a(p_2)\tilde{\phi}_b(p_3)\tilde{\phi}_b(p_4)~(2\pi)^d\delta^d(p_1+...+p_4)~\tilde{V}_{\rm eff}(p_1,p_2,p_3,p_4).\nonumber\\
\end{eqnarray}
The definition of $\Delta \mu_{\rm eff}^2$ and $\tilde{V}_{\rm eff}$ are obvious. 
\subsection{The Noncommutative Wilson-Fisher Fixed Point}
Next, we apply the Wilson renormalization group recursion formula to get an explicit expression of this action.  We will assume for simplicity that $u_0=v_0$. After some calculation we obtain
\begin{eqnarray}
S_{\rm quad}%=\int_{p\leq \rho\Lambda}\tilde{\phi}_a(p)\bigg(p^2+\mu^2+\Delta \Gamma_2(p)\bigg)\tilde{\phi}_a(-p)
&=&\Lambda^{d+2}\int_{{p}\leq \rho}\tilde{\phi}_a(\Lambda {p})\bigg[{p}^2+\bar{\mu}^2+\Delta \mu^2({p})\bigg]\tilde{\phi}_a(-\Lambda {p}).
\label{quadraticS}
\end{eqnarray}
\begin{eqnarray}
S_{\rm int}%&=&\frac{u_0}{N}\int_{p_1\leq \rho\Lambda}...\int_{p_4\leq \rho\Lambda }\tilde{\phi}_a(p_1)\tilde{\phi}_a(p_2)\tilde{\phi}_b(p_3)\tilde{\phi}_b(p_4)~(2\pi)^d\delta^d(p_1+...+p_4)~\tilde{V}_{\rm eff}(p_1,..,p_4)\nonumber\\
&=&\frac{\bar{u}_0}{N}\Lambda^{2d+4}\int_{p_1\leq \rho}...\int_{p_4\leq \rho}\tilde{\phi}_a(\Lambda p_1)\tilde{\phi}_a(\Lambda p_2)\tilde{\phi}_b(\Lambda p_3)\tilde{\phi}_b(\Lambda p_4)~(2\pi)^d\delta^d(p_1+...+p_4)~\tilde{V}_{\rm eff}(\Lambda p_1,..,\Lambda p_4).\nonumber\\\label{quarticS}
\end{eqnarray}
$\bar{\mu}^2$, $\bar{u}_0$ and $\bar{\theta}$ are dimensionless parameters, viz $\bar{\mu}^2=\mu^2/\Lambda^2$, $\bar{u}_0=u_0/\Lambda^{\epsilon}$ and $\bar{\theta}=\theta\Lambda^2$. The corrections $\Delta\mu^2$ and $\tilde{V}_{\rm eff}$ are now given by (with $V_d=\int_k 1/\Lambda^d$)
\begin{eqnarray}
\Delta \mu^2({p})&=&2\frac{\bar{u}_0}{1+\bar{\mu}^2}V_d~C_2(a).
\end{eqnarray}
\begin{eqnarray}
\tilde{V}_{\rm eff}(\Lambda p_1,..,\Lambda p_4)&=&\bigg(1-4\bar{u}_0 V_d\Delta^2(\bar{\lambda}) C_4(a)\bigg)\cos\Lambda^2 \frac{p_1\wedge p_2}{2}\cos\Lambda^2 \frac{p_3\wedge p_4}{2}\nonumber\\
&+&\frac{1}{2}\bigg(\cos\Lambda^2\frac{p_1\wedge p_3+p_2\wedge p_4}{2}+\cos\Lambda^2\frac{p_1\wedge p_4+p_2\wedge p_3}{2}\bigg).
\end{eqnarray}
$C_2$ and $C_4$ are the connected two-point and four-point functions of the  zero-dimensional vector model which is given by the functions \cite{Hikami:1978ya,Nishigaki:1996ts}
\begin{eqnarray}
C_2(a)=\frac{\sqrt{1+4a}-1}{2a}=1-a+2a^2-5a^3+...
\end{eqnarray}
\begin{eqnarray}
C_4(a)=\frac{1-(1+4a)^{-\frac{1}{2}}}{2a}=1-3a+10a^2-35a^3+...
\end{eqnarray}
The effective coupling $a$ is defined by 
%In this case
\begin{eqnarray}
a=\frac{V_d\bar{u_0}}{(1+\bar{\mu}^2)^2}~\bigg(1+\tilde{X}_{d-1}(\bar{\theta})\bigg)~,~
\tilde{X}_{d-1}(x)&=&\bigg(\frac{2}{x}\bigg)^{\frac{d}{2}-1}\Gamma\bigg(\frac{d}{2}\bigg)J_{\frac{d-2}{2}}(x).\label{a}
\end{eqnarray}
The renormalization group equations which follow from the above effective action are given by the equations

%\begin{eqnarray}
%\hat{\mu}^2&=&\rho^{-2}\bigg(\bar{\mu}^2+(1+\bar{\mu}^2)\frac{(1+4a)^{\frac{1}{2}}-1}{1+\tilde{X}_{d-1}(\bar{\theta})}\bigg).
%\end{eqnarray}
\begin{eqnarray}
\bar{\mu}^{'2}&=&\rho^{-2}\bigg(\bar{\mu}^2+t(1+\bar{\mu}^2)\frac{(1+4a)^{\frac{1}{2}}-1}{2}\bigg).
\end{eqnarray}
%\begin{eqnarray}
%\hat{u}_0+\hat{u}_0^{'}&=&2\rho^{-\epsilon}\bar{u}_0\bigg(1-\frac{1-(1+4a)^{-\frac{1}{2}}}{1+\tilde{X}_{d-1}(\bar{\theta})}\bigg).
%\end{eqnarray}
\begin{eqnarray}
\bar{u}_0^{'}+\bar{v}_0^{'}&=&2\rho^{-\epsilon}\bar{u}_0\bigg(1-t\frac{1-(1+4a)^{-\frac{1}{2}}}{2}\bigg).
\end{eqnarray}
%\begin{eqnarray}
%\hat{u}_0-\hat{u}_0^{'}=-2\rho^{-\epsilon}\bar{u}_0\frac{1-(1+4a)^{-\frac{1}{2}}}{1+\tilde{X}_{d-1}(\bar{\theta})}.
%\end{eqnarray}
\begin{eqnarray}
\bar{u}_0^{'}-\bar{v}_0^{'}=-\rho^{-\epsilon}\bar{u}_0t(1-(1+4a)^{-\frac{1}{2}}).
\end{eqnarray}
We have set
\begin{eqnarray}
t=\frac{2}{1+\tilde{X}_{d-1}(\bar{\theta})}=1+\frac{\bar{\theta}^2}{4d}+\frac{\bar{\theta}^4}{8d^2(d+2)}+....
\end{eqnarray}
We will also use the notation
\begin{eqnarray}
T=t\frac{(1+4a)^{\frac{1}{2}}-1}{2}.
\end{eqnarray}
A non-gaussian fixed point is given by $\bar{\mu}^{'2}=\bar{\mu}^2=\mu_*^2$ and $\bar{u}_0^{'}+\bar{v}_0^{'}=2\bar{u}_0=2u_*$ or equivalently
\begin{eqnarray}
\mu_*^2=\frac{T_*}{\rho^2-1-T_*}.
\end{eqnarray}

\begin{eqnarray}
\rho^{\epsilon}&=&1-t\frac{1-(1+4a_*)^{-\frac{1}{2}}}{2}.\label{contr1}
\end{eqnarray}
The fixed point near $\epsilon\simeq 0$, for any value of the dilatation parameter $\rho$, is given by
\begin{eqnarray}
a_*=-\frac{\epsilon\ln\rho}{t}~,~\mu_*^2=-\frac{\epsilon\ln\rho}{\rho^2-1}~,~u_*=-\frac{\epsilon\ln\rho}{2tV_d}.
\end{eqnarray}
For  $0<\epsilon\leq 2$, the critical values of $a$ and $T$ are given by
\begin{eqnarray}
 a_*=f(z)=\frac{(1-z)(t-1+z)}{(t-2+2z)^2}~,~z=\rho^{\epsilon}.\label{eqnt}
\end{eqnarray}
%We have in this case on the perturbative sheet $1-t/2\leq\rho^{\epsilon}\leq 1$ we have 
\begin{eqnarray}
T_*=t\frac{1-\rho^{\epsilon}}{t-2+2\rho^{\epsilon}}.
\end{eqnarray}
We must clearly have $\rho^{\epsilon}\geq 1-t/2$ and $\rho^{\epsilon}\leq 1$ for $T_*$ to be positive definite. We get therefore
\begin{eqnarray}
\mu_*^2&=&t\frac{1-\rho^{\epsilon}}{2(1-t)+(t-2)(\rho^2+\rho^{\epsilon})+2\rho^{\epsilon+2}}~,~\nonumber\\
u_*&=&\frac{t}{2V_d}\frac{(1-\rho^2)^2(1-\rho^{\epsilon})(t-1+\rho^{\epsilon})}{\big(2(1-t)+(t-2)(\rho^2+\rho^{\epsilon})+2\rho^{\epsilon+2}\big)^2}. \label{fixedP}
\end{eqnarray}
We can check that we have always  $u_*\geq 0$ and  $\mu_*^2\leq 0$.

The perturbative solution (\ref{contr1}) works on the perturbative sheet $1-t/2\leq \rho^{\epsilon}\leq 1$. However, there should be no difference between the regions $1-t/2\leq \rho^{\epsilon}\leq 1$ and  $0\leq \rho^{\epsilon}\leq 1-t/2$ and thus one must analytically continue the above solution to the non-perturbative sheet  $0\leq \rho^{\epsilon}\leq 1-t/2$.

On the perturbative sheet the function $f(z)$ starts from $f=0$ at $z=1$ then increases to $\infty$ as $z$ decreases to $z=1-{t}/{2}$, whereas on the non-perturbative sheet it starts from $f=(t-1)/(t-2)^2>0$ at $z=0$ then increases to $\infty$ as $z$ increases to $z=1-{t}/{2}$. The function $f(z)$ vansihes at $z=1-t<0$. We have then
\begin{eqnarray}
\sqrt{1+4a_*}=\pm\frac{t}{t-2+2z}.\label{sheets}
\end{eqnarray}
The plus sign corresponds to the above perturbative solution.  The minus sign in (\ref{sheets}) leads to
\begin{eqnarray}
2-t-\rho^{\epsilon}&=&1-t\frac{1-(1+4a_*)^{-\frac{1}{2}}}{2}.\label{contr2}
\end{eqnarray}
By setting $z=2-t-\rho^{\epsilon}$ on the second sheet, equation (\ref{contr2}) becomes equation  (\ref{contr1}).
%\begin{eqnarray}
%z&=&1-t\frac{1-(1+4a_*)^{-\frac{1}{2}}}{2}.
%\end{eqnarray}
This leads to $ a_*=f(z)$ where $f(z)$ is given by equation (\ref{eqnt}). The first sheet corresponds to the interval $z\in[1-t/2,1]$ whereas the second sheet corresponds to $z\in[0,1-t/2]$. When we continue the solution to the second sheet we observe that the critical coupling constant $a_*$ does not return to $0$ when we take the limit $\rho^{\epsilon}\longrightarrow 0$. Indeed, $f(0)=(t-1)/(t-2)^2>0$. As long as $\theta$ is sufficiently small we have $t$ near $1$ and as a consequence $f(0)$ is small and we get the commutative result.  The critical value of $T$ on the second sheet is
 \begin{eqnarray}
T_*&=&-\frac{t(t-1+\rho^{\epsilon})}{t-2+2\rho^{\epsilon}}.
\end{eqnarray}
As a consequence the critical values $\mu_*^2$ and $u_*$ on the second sheet are given by
\begin{eqnarray}
\mu_*^2&=&-\frac{t(t-1+\rho^{\epsilon})}{2-2t+t^2+(t-2)(\rho^2+\rho^{\epsilon})+2\rho^{2+\epsilon}}~,~\nonumber\\
u_*&=&\frac{t}{2V_d}\frac{(t-1+\rho^{\epsilon})(1-\rho^{\epsilon})(1-\rho^2)^2}{\big(2-2t+t^2+(t-2)(\rho^2+\rho^{\epsilon})+2\rho^{2+\epsilon}\big)^2}.\label{fixedNP}
\end{eqnarray}
This is the noncommutative Wilson-Fisher fixed point. 

We observe that for a fixed $t$ the limit of the perturbative fixed point (\ref{fixedP}) when $\rho^{\epsilon}\longrightarrow 1$ is (with $1/\tilde{V}_d=2^{d-1}\pi^{d/2}\Gamma(d/2)$)
\begin{eqnarray}
a_*=-\frac{\epsilon\ln\rho}{t}~,~\mu_*^2=-\frac{\epsilon}{2}~,~u_*=\frac{\epsilon}{2\tilde{V}_d}.
\end{eqnarray}
The limit for a fixed $t$ of the non-commutative Wilson-Fisher fixed point (\ref{fixedNP}) when $\rho^{\epsilon}\longrightarrow 0$ is
\begin{eqnarray}
a_*&=&\frac{t-1}{(t-2)^2}[1+\frac{\rho^{\epsilon}}{t-1}-\rho^{\epsilon}-\frac{4}{t-2}\rho^{\epsilon}+...]~,~\nonumber\\
\mu_*^2&=&-\frac{t(t-1)}{(t-1)^2+1}[1+\frac{\rho^{\epsilon}}{t-1}-\frac{t-2}{(t-1)^2+1}\rho^{\epsilon}+...]~,~\nonumber\\
u_*&=&\frac{1}{2V_d}\frac{t(t-1)}{((t-1)^2+1)^2}[1+\frac{\rho^{\epsilon}}{t-1}-\rho^{\epsilon}-2\frac{t-2}{(t-1)^2+1}\rho^{\epsilon}+...].
\end{eqnarray}
In contrast with the commutative theory and with the perturbative fixed point, the noncommutative Wilson-Fisher fixed point is not vanishingly small in the limit $\rho^{\epsilon}\longrightarrow 0$ and becomes significantly more important as we increase $t$ from $1$ to $2$, i.e. as we increase the noncommutativity $\bar{\theta}$ from $0$ to $\pi$. Indeed, we see that $a_*\longrightarrow \infty$ when $t\longrightarrow 2$, i.e. when we have only one sheet $[0,1]$. Putting it differently, in the limit $t\longrightarrow 2$ the two-sheeted structure of $a_*$ disappears and we end up only with the perturbative fixed point (\ref{fixedP}). 

% and as a consequence the fixed points at $\rho^{\epsilon}=0$ and $\rho^{\epsilon}=1$ seem to be different. 

%\begin{figure}[htbp]
%\begin{center}
%\includegraphics[width=4cm,angle=0]{vert1.eps}
%\includegraphics[width=10.0cm,angle=0]{vert2.eps}
%\caption{The vertex of non-commutative $O(N)$ sigma model.}\label{figure10}
%\end{center}
%\end{figure}
%\begin{figure}[htbp]
%\begin{center}
%\includegraphics[width=14.0cm,angle=0]{vert_dress1.eps}
%\includegraphics[width=10.0cm,angle=0]{one_loop_dressed.eps}
%\includegraphics[width=12.0cm,angle=0]{two_loop_dressed.eps}
%\includegraphics[width=8.0cm,angle=0]{three_loop_dressed.eps}
%\caption{The dressed vertex of non-commutative $O(N)$ sigma model.}\label{figure30}
%\end{center}
%\end{figure}

\end{document}